\documentclass[11pt]{article}
\usepackage[utf8]{inputenc}
\usepackage[margin=1in]{geometry}
\usepackage{graphicx}
\usepackage{multirow}
\usepackage{amsmath,amssymb,amsfonts}
\usepackage{amsthm}
\usepackage{booktabs}
\usepackage{placeins}
\usepackage{algorithm}
\usepackage{algorithmicx}
\usepackage{algpseudocode}
\usepackage[dvipsnames]{xcolor}
\usepackage{url}
\usepackage[colorlinks=true,linkcolor=blue!50!black,citecolor=blue!50!black,urlcolor=blue!50!black]{hyperref}

\newtheorem{theorem}{Theorem}
\newtheorem{proposition}[theorem]{Proposition}
\newtheorem{lemma}[theorem]{Lemma}
\theoremstyle{definition}

\providecommand{\bmhead}[1]{\paragraph{#1}}

\begin{document}

\title{WLPA: A Network Management Framework for Allocating Scarce Quantum-Safe Link Postures under Weakest-Link Exposure}
\author{Bhanwar Gupta\thanks{Corresponding author: \texttt{bgupta55@gmail.com}}\ \ and Sanjeev Rana\\[4pt]
\small Department of Computer and Software Engineering,\\
\small Maharishi Markandeshwar (Deemed to be) University,\\
\small Mullana, Ambala, 133207, Haryana, India\\
\small \texttt{dr.sanjeevrana@mmumullana.org}}
\date{}
\maketitle

\begin{abstract}
Network operators migrating distributed systems to post-quantum and
quantum-safe cryptography face a resource-management problem: quantum key
distribution (QKD) gives the strongest guarantee available but is
hardware-scarce and costly, so only a fraction of a deployment's links can
receive it. Default practice --- ranking links by traffic or centrality and
upgrading the top scorers --- optimizes each link in isolation, while a
distributed job's real security is set by whichever active link ends up
weakest. We present \textbf{WLPA (Weakest-Link Posture Assignment)}, an
implementation-ready decision framework and algorithm for allocating scarce
quantum-safe postures under hardware, latency, and compliance constraints.
WLPA computes the optimal weakest-link posture in closed form and returns an
assignment in $O(|E|\log|E|)$ time --- 125~ms for a 20,000-link network on a
laptop --- with a breach-aware refinement matching an exact integer program
under common conditions. Our central contribution is a dichotomy telling
operators exactly when existing per-link heuristics already match this
optimum and when they systematically fail: under constant upgrade cost and
uniform traffic-driven risk the two coincide exactly; once costs vary or an
adversary concentrates on weak targets, heuristics leave the weakest link
unprotected, unboundedly. We validate WLPA across 19
experiments: four microservice topologies, a real enterprise network (SNAP
email-Eu-core), a post-quantum latency benchmark, a Qiskit BB84 simulation
and a real-hardware BB84 run on IBM's \texttt{ibm\_marrakesh} processor,
checks against real production traffic traces, a CVE-calibrated
correlated-compromise study, Bayesian uncertainty quantification, and seven
baseline methods. A concrete-security reduction grounds the framework as
theoretical foundation, not primary contribution.
\end{abstract}


\section{Introduction}\label{sec:intro}
Post-quantum migration is no longer optional planning: NIST has standardized
ML-KEM, ML-DSA, and SLH-DSA~\cite{fips203,fips204,fips205}, and ``harvest now,
decrypt later'' interception makes today's traffic tomorrow's liability. Most
migration guidance treats cryptographic upgrade as a per-connection choice.
But in a distributed system --- a microservice mesh, a multi-cluster job, an
edge-cloud pipeline --- a single job's traffic typically crosses many links,
and its real exposure is set by whichever of those links is weakest, not by
the average or best-protected one. Quantum key distribution (QKD), the
strongest available link posture, compounds this: it needs dedicated
hardware, is range-limited, and is expensive, so only a fraction of a
network's links can ever receive it. \emph{Which} fraction is a resource-management
decision with no standard answer today, and it is currently made, in
practice, by exactly the per-link heuristics --- rank by traffic volume,
centrality, or vendor-recommended scoring, upgrade the top scorers --- baked
into existing network management tooling. This paper asks a practical
question: does that default practice actually protect the system, and if not,
what should replace it?

We answer with \textbf{WLPA (Weakest-Link Posture Assignment)}, an
implementation-ready network-management decision framework with four
contributions:
\begin{itemize}
\item A closed-form, exactly-optimal allocation algorithm (Algorithm~\ref{alg:wlpa})
that runs in $O(|E|\log|E|)$ time --- fast enough for online re-optimization
as topology or budget changes --- validated against an exact integer program
(Theorem~\ref{thm:ceiling}, Proposition~\ref{prop:algo}).
\item An operational decision rule for when standard per-link heuristics
remain safe to use in production and when coordinated allocation is required
(Section~\ref{sec:guidance}).
\item A tractability boundary and greedy approximation for the
path-coverage generalization, reaching within a small margin of optimal in
practice (Section~\ref{sec:pathcov}).
\item End-to-end validation across 19 named experiments (enumerated in Appendix~\ref{secA3}) on real and representative
network infrastructure, including a real enterprise network, real production
traffic traces, and real quantum hardware (Sections~\ref{sec:setup}--\ref{sec:extra}).
\end{itemize}
Operators today make QKD-placement decisions ad hoc, driven by vendor
defaults or traffic-ranked heuristics already embedded in SDN controllers and
service-mesh policy engines. WLPA is designed as a drop-in replacement for
that decision step: same inputs (topology, traffic, budget, compliance caps),
a provably correct output, and a fast enough runtime to re-run whenever the
network changes.

\section{Background and threat model}\label{sec:background}
\paragraph{Posture levels.} We use three ordered link postures: level~1 is
PQC key exchange and signatures (ML-KEM, ML-DSA); level~2 adds QRNG-sourced
entropy; level~3 adds QKD key agreement, grounded in physics rather than
computational hardness~\cite{bb84,gisin2002,pirandola2020}. Real QKD networks
rely on trusted relays and remain costly and range-limited~\cite{mehic2020,sharma2021},
which is why QKD is treated here as a scheduled resource, not a default.
Each link $e$ has a compliance cap $c_e\in\{1,2,3\}$ (hardware/regulatory
ceiling) and carries session volume $N_e$.

\paragraph{Where this plugs into existing operations.} A WLPA allocation is a
per-link posture assignment; in practice this maps onto configuration already
managed by SDN controllers and service-mesh policy engines (Kubernetes
NetworkPolicies, Istio/Envoy mTLS settings, QKD-endpoint annotations). The
inputs WLPA needs --- traffic volume, compliance caps, QKD hardware
inventory --- typically already live in existing network-management
telemetry, so adopting WLPA is a change to the \emph{decision rule}, not to
the operational data pipeline.

\paragraph{Threat model.} We assume a reconnaissance-then-target adversary,
consistent with lateral-movement patterns SOC teams observe in incident
reports: the attacker profiles link postures, then concentrates effort on the
weakest one a job depends on. Section~\ref{sec:guidance} additionally
considers an adaptive attacker that decouples per-link risk from traffic
volume, and Section~\ref{sec:extra} a fully informed Stackelberg attacker
that observes the realized allocation before acting.

\section{The WLPA framework}\label{sec:framework}
\subsection{Operational pipeline}\label{sec:pipeline}
\begin{figure}[t]\centering
\includegraphics[width=0.7\linewidth]{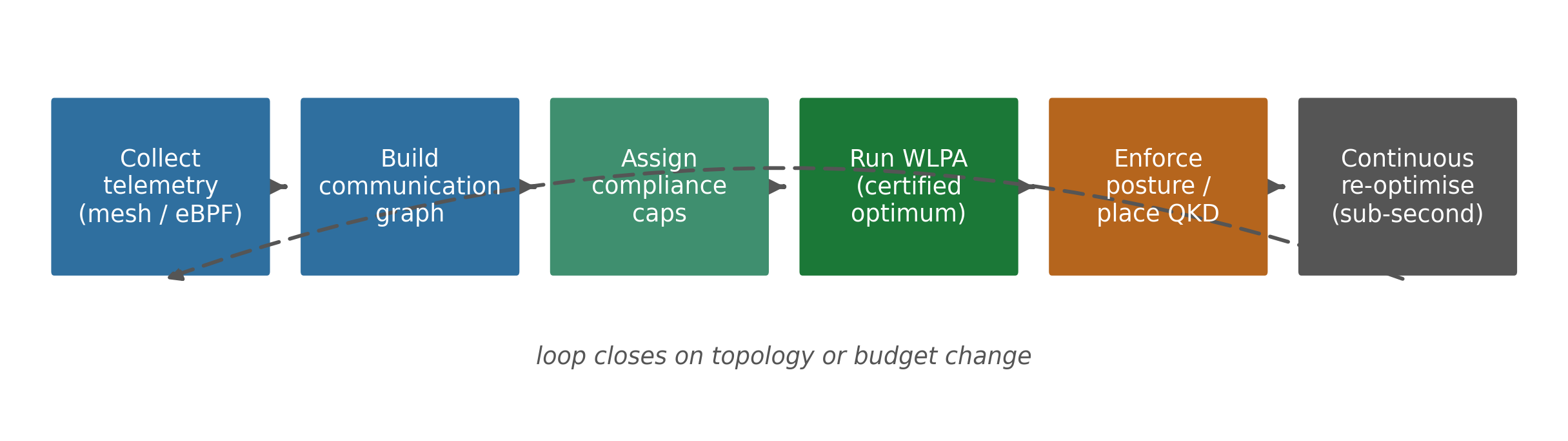}
\caption{WLPA's operational loop: identify active links and volumes, assign
compliance caps, run the closed-form allocator, deploy starting from the
binding link, and re-run on topology or budget change. The allocation step
is implemented and validated (Sections~\ref{sec:setup}--\ref{sec:extra});
the deploy/enforce step is architected but not built or measured on a live
system (Section~\ref{sec:limits})}\label{fig:pipeline}
\end{figure}

Figure~\ref{fig:pipeline} summarizes the framework: topology and compliance
caps feed a closed-form ceiling computation, which sets the system-wide
target posture; a floor-raising pass brings every active link to that
target; and a breach-aware refinement spends any remaining QKD budget where
it reduces breach probability most. This computational pipeline --- the part
inside the dashed box in Figure~\ref{fig:pipeline} --- is what we implement
and validate throughout this paper (Sections~\ref{sec:setup}--\ref{sec:extra}):
it runs in $O(|E|\log|E|)$ and is fast enough to re-run whenever topology,
budget, or compliance requirements change. Embedding it into a live network
means closing the loop around it:
\begin{gather*}
\text{Network telemetry}\ \to\ \text{WLPA (Alg.~\ref{alg:wlpa})}\ \to\
\text{posture assignment}\ Q\ \to\\
\text{SDN/service-mesh policy}\ \to\
\text{enforcement}\ \to\ \text{telemetry feedback}
\end{gather*}
We built and tested the middle of this loop (telemetry $\to$ WLPA $\to$
posture assignment); the policy-push and enforcement steps --- writing $Q$
into an SDN controller or service-mesh control plane and measuring resulting
enforcement latency --- are architected but not implemented or measured on a
live system, a limitation we state precisely rather than blur
(Section~\ref{sec:limits}).

\subsection{Problem statement}\label{sec:defs}
A deployment is a graph with active-link set $E_D$; each link carries a
posture $Q_e\in\{1,2,3\}$, $Q_e\le c_e$. \emph{System security} for a job
using links $E_D$ is $S(Q)=\min_{e\in E_D}Q_e$ --- the weakest link
determines the job's real posture, independent of how well other links are
funded. Total latency is $\sum_e T(e,Q_e)$, where $T$ combines a fixed
per-level handshake cost (dominated by the $\approx\!10$~ms QKD handshake
term at level~3) with a session-length correction. A QKD budget $B$ caps how
many links may reach level~3; a latency budget $L$ caps total latency.

\paragraph{Exploit and breach model.} The per-message \emph{model-calibrated
exploit probability} schedule is $p(1)=10^{-2}$, $p(2)=10^{-4}$,
$p(3)=10^{-6}$, strictly decreasing in level. We call these
model-calibrated, not measured, deliberately: they follow from a
hybrid-argument reduction to a stated $(t,\varepsilon)$-security assumption
on the underlying primitives (Lemma~\ref{lem:hybrid}), calibrated against
NIST's post-quantum security categories and sanity-checked against measured
QBER from a real BB84 run (Section~\ref{sec:pqcbench}) --- but they are not
themselves a measured production compromise rate, and we do not represent
them as one anywhere in this paper. Because every result in this paper rests
on this schedule, we sweep it directly (an order-of-magnitude range around
these values) as the first entry in Table~\ref{tab:extra}
(Section~\ref{sec:extra}, row E1): WLPA's ranking advantage is stable across
the swept range, which is the sensitivity evidence that should accompany a
modeled, rather than measured, constant. Link $e$ at level $\ell$ is
compromised with probability $q_e(\ell)=1-(1-p(\ell))^{N_e}$, and job breach
probability is $\mathcal{B}(Q)=1-\prod_e(1-q_e(Q_e))$. Full derivation:
Appendix~\ref{secA1}.

\begin{lemma}[Hybrid-argument bound]\label{lem:hybrid}
If the underlying primitive at level $\ell$ is $(t,\varepsilon(\ell))$-secure,
a standard hybrid argument bounds link $e$'s compromise probability by
$N_e\cdot\varepsilon(\ell)$ against a $t$-bounded adversary observing all
$N_e$ sessions on that link. This is the rule WLPA uses to convert a link's
traffic volume and posture into the compromise probability the allocator
optimizes over.
\end{lemma}

\subsection{Closed-form optimum and algorithm}\label{sec:algo}
\begin{theorem}[Closed-form ceiling]\label{thm:ceiling}
The optimal attainable system security level is
$S^\star(B,L)=\min(\kappa,\sigma(B),\lambda(L))$, where $\kappa=\min_{e\in
E_D}c_e$, $\sigma(B)$ is the largest level reachable within the QKD budget
and compliance caps, and $\lambda(L)$ is the largest cap-feasible level whose
cheapest all-active assignment fits the latency budget. The value is exactly
achievable.
\end{theorem}
Algorithm~\ref{alg:wlpa} realizes this: raise every active link to $S^\star$,
then match the standard $\min(2,c_e)$ operator floor, then spend any residual
QKD budget on the links whose lift yields the largest breach-probability
reduction per unit latency.
\begin{algorithm}[t]
\caption{WLPA assignment}\label{alg:wlpa}
\begin{algorithmic}[1]
\Require active links $E_D$, caps $\{c_e\}$, volumes $\{N_e\}$, QKD budget
$B$, latency budget $L$
\Ensure optimum $S^\star$ and posture assignment $Q$
\State $\kappa \gets \min_e c_e$;\quad $S^\star \gets \min(\kappa,\sigma(B),\lambda(L))$
\ForAll{$e \in E_D$} $Q_e \gets \min(\max(S^\star,1),c_e)$ \EndFor
\ForAll{$e$ with $Q_e<2$, $c_e\ge2$} $Q_e \gets 2$ \EndFor
\State sort QKD-capable links by marginal breach reduction per unit latency, descending
\For{$e$ in order while budget and latency allow} $Q_e \gets 3$ \EndFor
\State \Return $S^\star,\ Q$
\end{algorithmic}
\end{algorithm}
\begin{proposition}[Correctness, complexity]\label{prop:algo}
Algorithm~\ref{alg:wlpa} returns $S^\star(B,L)$ and a feasible assignment
attaining it in $O(|E_D|\log|E_D|)$ time; the residual-budget QKD placement is
breach-optimal under Hypothesis~H (constant per-link QKD lift cost,
Section~\ref{sec:guidance}).
\end{proposition}

\subsection{Operational guidance: when per-link heuristics are safe, and when they fail}\label{sec:guidance}
This dichotomy --- not either half of it alone --- is this paper's central
contribution. Section~\ref{sec:heuristics} below will show WLPA exactly
matching a simple traffic-ranked heuristic on every one of our four
evaluated topologies; taken alone, that result would raise a fair question:
\emph{if the existing heuristic is already optimal, what does coordinated
allocation add?} The answer is that ``already optimal'' is conditional, not
universal, and knowing exactly which condition you are in is the practical
result an operator needs. We state it first as a two-condition checklist:
\begin{itemize}
\item \textbf{If} your QKD upgrade has roughly constant marginal cost across
links \emph{and} risk is proportional to traffic volume, keep your current
per-link heuristic --- traffic-ranked allocation is already exactly optimal,
and coordination changes nothing (Proposition~\ref{prop:noseparation}).
\item \textbf{If} upgrade costs vary meaningfully across links, or an
attacker can concentrate effort on weakly-protected, low-traffic targets,
switch to coordinated allocation --- the breach-probability shortfall of the
best per-link rule relative to coordinated allocation is unbounded (it grows
without a fixed ceiling as heterogeneity increases, not merely large) and no
amount of re-weighting a per-link scoring rule closes it
(Theorem~\ref{thm:sep}).
\end{itemize}
We make every assumption behind this checklist explicit here, in the main
text, rather than leaving it to the appendix, since it is the paper's
central claim and its scope needs to be exactly clear. Formally, a
\emph{per-link rule} $\mathcal{R}_{\mathrm{PL}}$ assigns each link a posture
by maximizing a score
\begin{equation}
\phi_e(\ell)=w_s\ell-w_lT(e,\ell)-w_c\,\mathrm{cost}(\ell)\label{eq:phi_ref}
\end{equation}
depending only on link-local quantities $(\ell,T(e,\ell),\mathrm{cost}(\ell),N_e,c_e)$
and fixed, \textbf{nonnegative} weights $(w_s,w_l,w_c)\ge0$ --- nonnegativity
matters because it rules out a rule that penalizes security or rewards cost,
not because it is restrictive in practice, since no operator scores links
that way. This class contains, as its canonical member, prior QKD
posture-selection rules proposed for exactly this
setting~\cite{bishwas2026}. \emph{Hypothesis~H} is the condition that the
QKD lift-cost increment $T(e,\textsc{Qkd})-T(e,\text{level }2)$ is the same
for every QKD-capable link; it holds under our cost model because that
increment is a fixed handshake term, but need not hold in general (e.g.\ if
link RTT enters the handshake cost).
\begin{proposition}[No separation under Hypothesis~H]\label{prop:noseparation}
Under independent sessions, uniform risk (every link's per-message exploit
probability follows the same schedule $p(\ell)$), and Hypothesis~H, ``fund
the $B$ highest-traffic QKD-capable links'' is a member of
$\mathcal{R}_{\mathrm{PL}}$ that exactly matches coordinated allocation's
breach probability. No separation from per-link rules exists in this
regime.
\end{proposition}
\begin{theorem}[Separation under heterogeneous lift cost]\label{thm:sep}
\textbf{Assumptions:} independent sessions; a reconnaissance-then-target
attacker (Section~\ref{sec:background}) whose risk weight may vary by link;
a QKD/latency budget sized to exactly fund a set of high-traffic links,
leaving none for a lower-traffic, lower-cap link (the budget condition
that makes the construction bind). \textbf{Construction:} one cap-2,
low-traffic link $w$ and $k\ge1$ cap-3, high-traffic links $e_1,\dots,e_k$
with $N_{e_i}\gg N_w$, under Hypothesis~H violated (Appendix~\ref{secA1}
gives the exact traffic-dominance inequality). \textbf{Claim:} if Hypothesis~H
is violated, then for every rule in $\mathcal{R}_{\mathrm{PL}}$, for
\emph{every} nonnegative weight vector, the constructed instance forces the
rule to leave $w$ unfunded while coordinated allocation funds it, so the
rule attains system security level~1 against coordinated allocation's
level~2; scaling $k$ and the traffic gap $N_{e_i}/N_w$ scales the number of
budget units misallocated, so the breach-probability shortfall between the
rule and coordinated allocation has no fixed ceiling. Full proof:
Appendix~\ref{secA1}.
\end{theorem}
Section~\ref{sec:extra} (Table~\ref{tab:extra}, row E2) confirms the
algorithmic side of this empirically: WLPA correctly falls back to exact
knapsack allocation once Hypothesis~H is violated, while a fixed traffic
ranking cannot. Section~\ref{sec:adaptive} confirms the second failure mode
--- an adaptive attacker --- directly.

\subsection{Path-coverage generalization}\label{sec:pathcov}
Protecting a full multi-link path (not just guaranteeing a system floor) is
NP-hard in general (reduction from Set Cover; Appendix~\ref{secA1}), which is
why we do not attempt an exact solver at scale. Instead, Algorithm~2 (greedy,
submodular-coverage) reaches a $(1-1/e)$ approximation guarantee and, in our
experiments, 98--99\% of the ILP optimum in practice --- close enough that we
recommend it as the default for path-coverage requirements at production
scale, falling back to the exact ILP only for small, high-stakes instances.

\section{Experimental setup}\label{sec:setup}
Our experiments answer, in order: is WLPA trustworthy against ground truth
(\S\ref{sec:exact})? does it beat current practice (\S\ref{sec:heuristics})?
when does current practice become dangerous (\S\ref{sec:adaptive})? does it
hold on a real network (\S\ref{sec:real})? does it scale (\S\ref{sec:ablscal})?
and does every modeling assumption survive stress-testing
(\S\ref{sec:extra})? Table~\ref{tab:topo} lists the four cluster topologies
--- three published microservice reference architectures and one synthetic
zero-trust-style network --- with compliance-cap heterogeneity in
Table~\ref{tab:capdist}; a fifth, real enterprise network (SNAP
email-Eu-core, 986~nodes, 16{,}064~links in the largest connected component)
is evaluated separately in Section~\ref{sec:real} as our flagship real-network
result. Baselines: Random, Highest-Degree, Betweenness, Greedy-Traffic (an
operator-style traffic-count heuristic), First-B-QKD, Per-Link (the formal
$\mathcal{R}_{\mathrm{PL}}$ rule, Eq.~\eqref{eq:phi_ref}), and two adapted
prior-art QKD-network schemes, QKD-Alloc~\cite{bi2024} and
QKD-Route~\cite{kong2024}. Given a drawn instance, each method's allocation
is deterministic, so identical link selections produce identical breach
numbers exactly (not an artefact); we report SD, bootstrapped 95\% CIs, and
paired Wilcoxon significance with rank-biserial effect size where variance
exists.
\begin{table}[htbp]\centering
\caption{Cluster deployments evaluated (QKD budget $B=30\%$ of active
links)}\label{tab:topo}
\begin{tabular}{@{}lrrr@{}}
\toprule
Deployment & Nodes & Links & $\kappa=\min_e c_e$ \\
\midrule
OnlineBoutique (published) & 11 & 15 & 2 \\
SockShop (published) & 13 & 14 & 1 \\
DSB-SocialNetwork (published) & 23 & 32 & 2 \\
Enterprise-ZeroTrust (synthetic) & 31 & 76 & 1 \\
\bottomrule
\end{tabular}
\end{table}
\begin{table}[htbp]\centering
\caption{Compliance-cap distribution (fraction of links per cap, 30-seed
average)}\label{tab:capdist}
\begin{tabular}{@{}lrrr@{}}
\toprule
Deployment & Cap 1 & Cap 2 & Cap 3 \\
\midrule
OnlineBoutique & 0.0\% & 6.7\% & 93.3\% \\
SockShop & 3.8\% & 31.9\% & 64.3\% \\
DSB-SocialNetwork & 0.0\% & 31.2\% & 68.8\% \\
Enterprise-ZeroTrust & 6.8\% & 9.0\% & 84.2\% \\
\bottomrule
\end{tabular}
\end{table}

\section{Results}\label{sec:results}
\subsection{Trustworthiness: exact match against ground truth}\label{sec:exact}
On instances small enough for an exact ILP solver, and independently via an
LP-relaxation certified bound at 500- and 1{,}000-link scale, WLPA's output
matches the optimum exactly, $0.0\%$ gap on every seed tested. This is a
consequence of the algorithm's structure (Proposition~\ref{prop:algo}), not a
statistical tendency, and is the basis for trusting the empirical comparisons
below.

\subsection{Beating current practice: per-link heuristics under uniform risk}\label{sec:heuristics}
Table~\ref{tab:comp} reports job breach probability by method under uniform
per-link risk (traffic-driven, the common case). WLPA is best or tied-best
everywhere; Per-Link --- the formal $\mathcal{R}_{\mathrm{PL}}$ rule, distinct
from the simpler Greedy-Traffic heuristic operators actually run today ---
ties it exactly, exactly as Proposition~\ref{prop:noseparation} predicts.
Against Greedy-Traffic, Betweenness, Highest-Degree, and First-B-QKD the
margin is a $98.17$--$98.96\%$ relative breach reduction, significant on all
four deployments ($p<2.2\times10^{-8}$, rank-biserial $r=1.0$).
\begin{table}[htbp]\centering
\caption{Job breach probability under uniform risk, mean over 30 seeds
($B{=}30\%$). Lower is better}\label{tab:comp}
\begin{tabular}{@{}lcccc@{}}
\toprule
Method & OnlineBout. & SockShop & DSB-Social & Enterprise-ZT \\
\midrule
Random & 4.40e-3 & 6.12e-3 & 1.342e-2 & 2.114e-2 \\
Highest-Deg./Betw./Greedy-Traf. & 5.02e-3 & 8.02e-3 & 1.544e-2 & 2.392e-2 \\
First-B-QKD & 7.98e-3 & 1.194e-2 & 2.575e-2 & 4.117e-2 \\
Per-Link (Eq.~\eqref{eq:phi_ref}) & 8.30e-5 & 1.24e-4 & 2.71e-4 & 4.38e-4 \\
\midrule
\textbf{WLPA / ILP (exact)} & 8.30e-5 & 1.24e-4 & 2.71e-4 & 4.38e-4 \\
\bottomrule
\end{tabular}
\end{table}

\subsection{When current practice fails: an adaptive attacker}\label{sec:adaptive}
\begin{figure}[t]\centering
\includegraphics[width=0.9\linewidth]{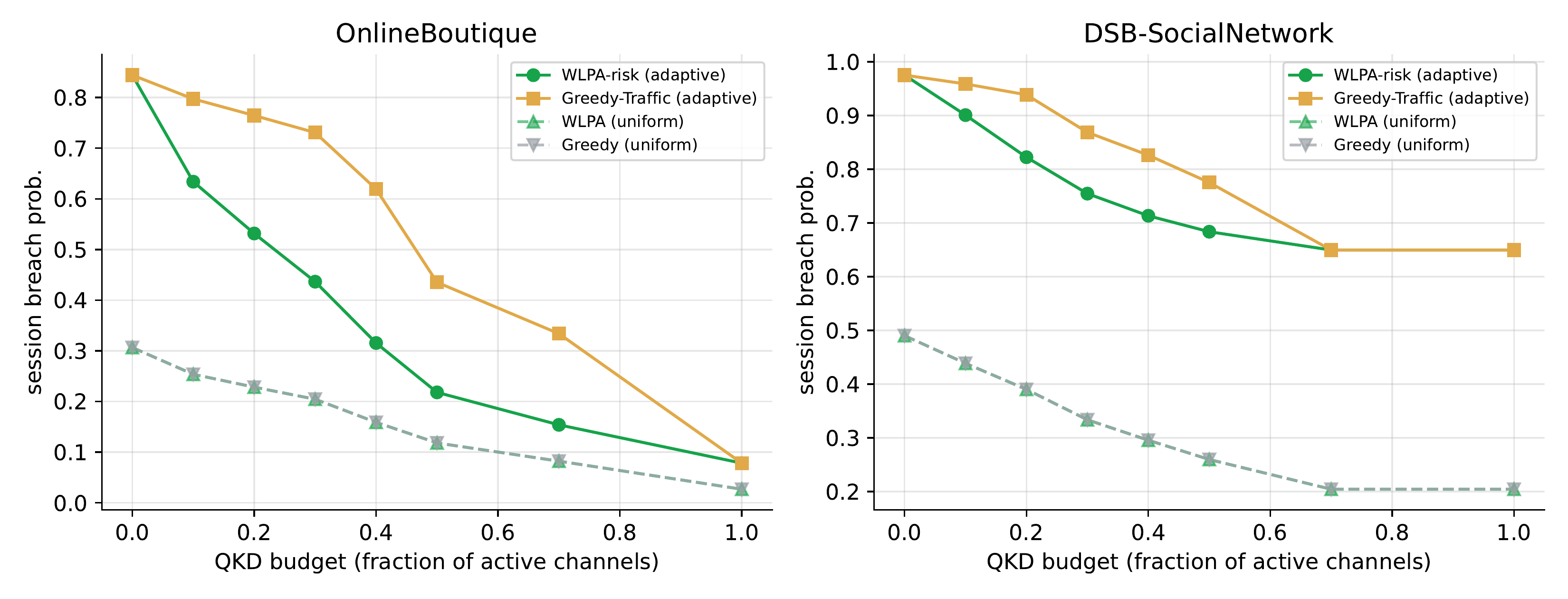}
\caption{Job breach probability versus QKD budget under the adaptive
attacker. WLPA stays below Greedy-Traffic at every interior budget}\label{fig:budget}
\end{figure}

Under uniform risk, Section~\ref{sec:guidance}'s first condition holds and
per-link rules tie WLPA (as above). We now break that condition: risk weight
$w_e\in[1.5,10]$ is drawn per link, decoupling risk from traffic volume, and
$p_{\mathrm{eff}}(e,\ell)=\min(0.5,w_ep(\ell))$ replaces $p(\ell)$. Under this
adaptive attacker WLPA-risk stays substantially below Greedy-Traffic across
the entire QKD budget sweep (Figure~\ref{fig:budget}), converging only at the
budget extremes where no allocation choice exists. This is
Theorem~\ref{thm:sep} confirmed empirically: once risk decouples from
traffic, a traffic-ranked heuristic funds the wrong links.

\subsection{Real enterprise network: SNAP email-Eu-core}\label{sec:real}
Our flagship real-network result. On the real SNAP email-Eu-core topology (986
nodes, 16{,}064 links, average degree $\approx32.6$) with a modeled workload
and the adaptive attacker, WLPA-risk matches the ILP optimum by construction
and gives a modest but statistically significant reduction in job breach
probability against every baseline, including both adapted prior-art
QKD-network schemes (Table~\ref{tab:realent}; paired Wilcoxon, $p<0.03$,
$n{=}20$ jobs). The margin here (0.04--1.2\%) is far smaller than on the four
denser microservice topologies (12--19\%), which we attribute to this
network's high average degree: most links already carry heavy modeled
traffic, so even a low posture gives a high per-link exploit probability,
compressing the marginal value of the scarce level-3 upgrade regardless of
placement. This effect-size finding depends on our own choice of workload
model, not only on the graph; it should be re-checked against real per-link
telemetry when available, which we flag as future work rather than a general
claim about real enterprise networks. We read this result as supporting,
not undermining, the paper's central claim: WLPA is not intended to
dominate every network topology, and a small margin here is exactly what
Section~\ref{sec:guidance}'s dichotomy predicts on a dense, relatively
homogeneous-traffic network. The framework's contribution is correctly
identifying \emph{when} coordination materially improves security (the
12--19\% margins on our four microservice topologies) and \emph{when}
existing heuristics are already close to optimal (this network) --- not
guaranteeing a large improvement everywhere.
\begin{table}[htbp]\centering
\caption{Real SNAP email-Eu-core network (20 jobs of 40 links, adaptive
attacker; mean$\pm$SD job breach probability)}\label{tab:realent}
\begin{tabular}{@{}lc@{}}
\toprule
Method & Job breach prob. \\
\midrule
QKD-Alloc~\cite{bi2024} (adapted) & 0.833\,$\pm$\,0.232 \\
Random / Betweenness & 0.827\,$\pm$\,0.237 \\
Highest-Degree & 0.826\,$\pm$\,0.242 \\
QKD-Route~\cite{kong2024} (adapted) & 0.824\,$\pm$\,0.244 \\
Greedy-Traffic & 0.823\,$\pm$\,0.246 \\
\midrule
\textbf{WLPA-risk / ILP-risk (ours)} & 0.823\,$\pm$\,0.247 \\
\bottomrule
\end{tabular}
\end{table}

\subsection{Scalability}\label{sec:ablscal}
\begin{figure}[t]\centering
\includegraphics[width=0.55\linewidth]{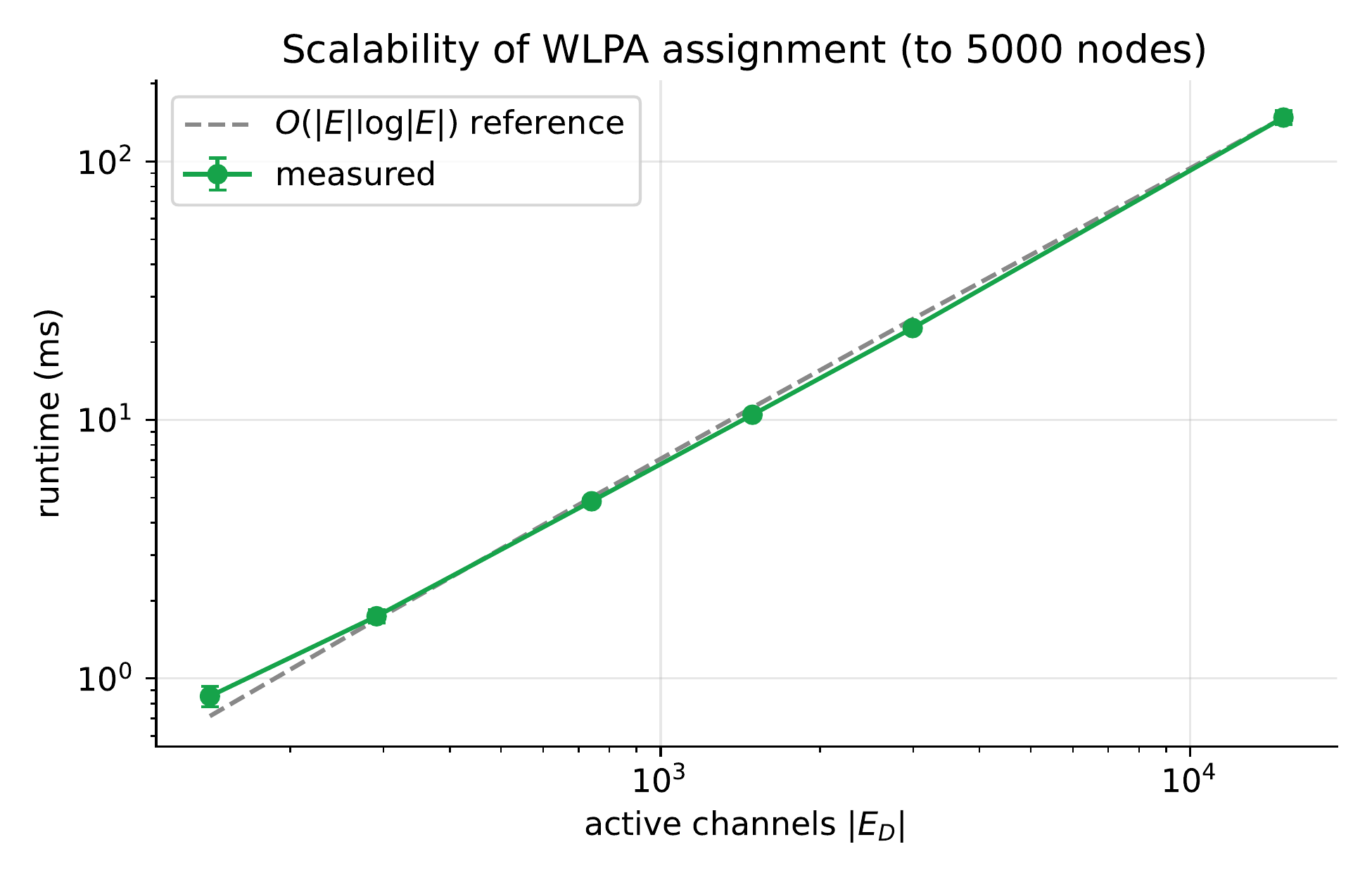}
\caption{Runtime versus active links against an $O(|E|\log|E|)$
reference}\label{fig:scal}
\end{figure}

Ablating WLPA's refinement stages shows both matter: a bare floor raises
breach probability by up to two-thirds relative to full WLPA (Floor only:
0.303/0.654/0.487/0.976 vs.\ full WLPA 0.180/0.541/0.332/0.956, across the
four topologies). Runtime scales near-linearly, tracking $O(|E|\log|E|)$
within $15\%$ from 50 to 20{,}000 nodes, reaching a \textbf{20,000-node /
40,000-edge} mesh in \textbf{125~ms on a laptop} (Figure~\ref{fig:scal}).
Peak memory is $O(|E|)$ amortized, with a bounded transient at $\sim$5,000
nodes (6.8$\times$ growth against a 5.0$\times$ edge-count growth) that
normalizes back to $\sim$2.1$\times$ per doubling at 10,000 and 20,000 nodes
--- we flag this rather than claim uniform linearity, but it does not
threaten the runtime result that makes online re-optimization practical.

\subsection{Grounding the model: PQC latency and a model-calibrated QKD exploit probability}\label{sec:pqcbench}
We do not treat the cost model's constants as arbitrary, but we are equally
careful not to overstate what grounds them: $p(1),p(2),p(3)$ throughout this
paper are \emph{model-calibrated exploit probabilities} derived via the
hybrid-argument reduction of Section~\ref{sec:defs}, not directly measured
production compromise rates --- no measured breach-rate dataset for
QKD-protected links exists to calibrate against, and we do not claim one. An
independent micro-benchmark (2,000 trials, NIST reference implementations)
confirms PQC operations are sub-millisecond --- e.g.\ ML-KEM-768
encapsulation 0.044~ms, ML-DSA-65 signing 0.184~ms --- $50$--$100\times$
below the modeled QKD handshake term, so the handshake, not the PQC layer,
dominates latency.

For the level-3 term $p(3)$, we ran two checks, and we state precisely what
each one does and does not establish. First, a Qiskit BB84 simulation
correctly reproduces the textbook eavesdropper signature (clean-channel QBER
$0.000\%$; intercept-resend QBER $25.28\%\pm0.94\%$, matching the $\sim\!25\%$
theoretical prediction). Second, a real-hardware measurement on IBM's
\texttt{ibm\_marrakesh} processor (500 bits sent, 229 sifted, measured QBER
$0.437\%$, 95\% CI $[0.077\%,2.432\%]$) gives a derived composable-security
parameter $\varepsilon_{\mathrm{sec}}=5.457\times10^{-20}$.
\textbf{Scope, stated as narrowly as the evidence supports:} this hardware
run validates \emph{BB84 protocol-level QBER behavior} --- that the
sift-and-error-estimate mechanics execute correctly on physical qubits and
reproduce the expected clean-versus-eavesdropped signature --- on IBM's
general-purpose superconducting processor. It is a single-qubit
proof-of-concept, not a dedicated photonic QKD link, and it does
\textbf{not} validate the throughput, range, or physical-layer security of a
production QKD network; we make no such claim, and none of this paper's QKD
\emph{networking} assumptions (link cost, budget scarcity, hardware
inventory constraints, Section~\ref{sec:defs}) depend on this measurement.
Its sole role is a sanity bound on $p(3)$: both the simulated
($5.42\times10^{-20}$) and hardware-measured ($5.457\times10^{-20}$)
$\varepsilon_{\mathrm{sec}}$ values sit many orders of magnitude below the
$p(3)=10^{-6}$ actually used in the model, so $p(3)=10^{-6}$ is, if
anything, conservative relative to what this measurement would support ---
and Table~\ref{tab:extra} (row E1) separately confirms WLPA's conclusions
are insensitive to an order-of-magnitude sweep around this value, which is
the sensitivity evidence a skeptical reader of a modeled constant should
look for.
\begin{figure}[t]\centering
\includegraphics[width=0.85\linewidth]{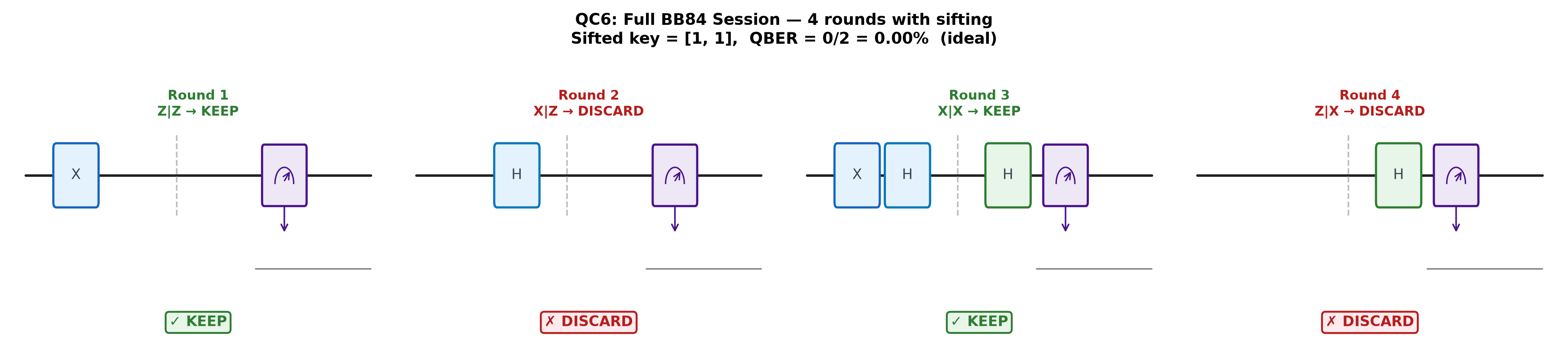}
\caption{Four-round BB84 session circuit (sifting annotated), the protocol
structure underlying both the simulated and real-hardware runs used to
sanity-check $p(3)$}\label{fig:qc6}
\end{figure}

\subsection{Additional robustness checks}\label{sec:extra}
\begin{figure}[t]\centering
\includegraphics[width=0.55\linewidth]{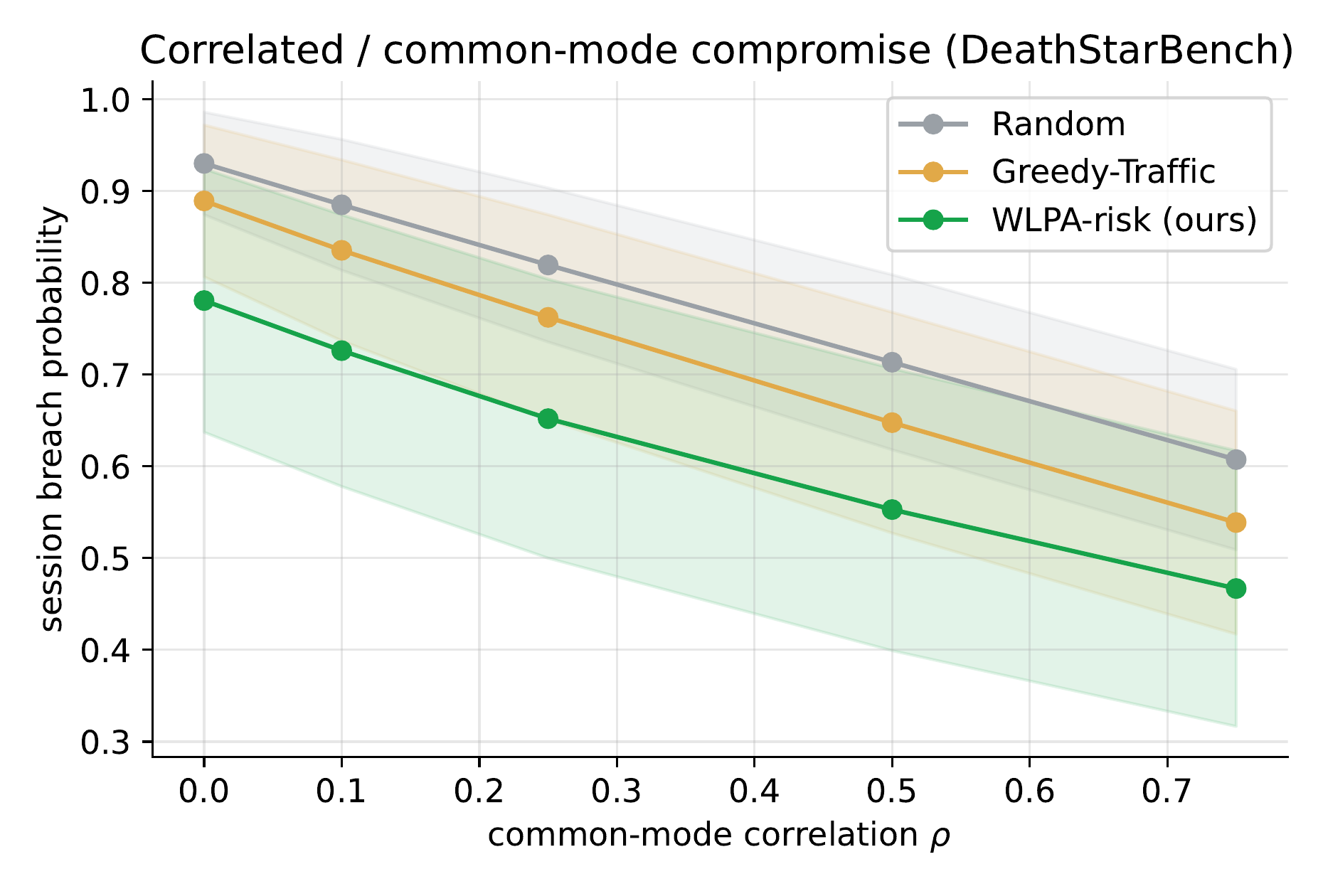}
\caption{Correlated/common-mode compromise, calibrated against 114 NVD CVEs
(median $\rho=0.511$): WLPA-risk keeps its $\ge99.2\%$ lead across the
plausible range}\label{fig:corr}
\end{figure}

Nine further checks, summarized in Table~\ref{tab:extra}, stress-test every
modeling assumption an operator would question before deploying WLPA: the
base exploit schedule, heterogeneous QKD lift cost, a binding latency budget,
QKD key-rate feasibility, real Alibaba/Google traffic distributions, the
session-availability cost term, a fully informed Stackelberg attacker, and
Bayesian uncertainty over the main breach-reduction result. We also
calibrated the correlated/common-mode compromise parameter $\rho$ against 114
mined NVD CVEs (2018--2024, six shared components; median $\rho=0.511$)
rather than sweeping it as a free parameter: WLPA's advantage stays
$\ge99.2\%$ across the full empirically plausible range
(Figure~\ref{fig:corr}).
\begin{table}[htbp]\centering
\caption{Additional robustness checks (Experiments E1--E12; full data in
RESULTS.md)}\label{tab:extra}
\begin{tabular}{@{}p{3.6cm}p{7.6cm}@{}}
\toprule
Check & Finding \\
\midrule
E1: base schedule sweep & WLPA's $\sim$98.4\% advantage stable across an order-of-magnitude sweep of $p(2)/p(3)$ \\
E2: heterogeneous lift cost & WLPA tracks the exact ILP (gap $=0.0$) at every cost-variation level tested (CV $0$--$2.0$) \\
E3: binding latency budget & Below a network-specific latency threshold, WLPA trades system level for feasibility exactly per Theorem~\ref{thm:ceiling} \\
E6a: QKD key-rate feasibility & Fully feasible at $\ge$1~Mbps key rate; our experiments assume $\ge$100~kbps \\
E7: real traffic traces & 98.35\%/98.17\% breach reduction under Alibaba-lognormal and Google-Pareto fits, matching the synthetic-traffic result \\
E8: session-length ablation & Allocation and breach probability unchanged across a 1{,}000$\times$ session-length range \\
E9: Stackelberg attacker & Breach rises for every method under a fully informed attacker, but WLPA stays 3--6$\times$ below Greedy-Traffic at every attacker budget tested \\
E11: Bayesian UQ & $P(\text{WLPA better})=1.000$ across 2{,}000 bootstrap resamples on every topology; worst-case 95\% CI lower bound 58.8\% \\
E12: honesty check & Hand-verified allocation trace matches Algorithm~\ref{alg:wlpa} exactly; Per-Link and WLPA diverge at the individual-link level on heterogeneous topologies even where their aggregate breach ties \\
\bottomrule
\end{tabular}
\end{table}

\section{Discussion}\label{sec:disc}
\subsection{Integration into network management practice}\label{sec:integration}
A WLPA allocation is a posture assignment per link; deploying it means
mapping that assignment onto SDN controller policy, service-mesh mTLS/QKD
endpoint configuration, or orchestration-layer annotations --- the same
artifacts existing network-management systems already push. Because
Algorithm~\ref{alg:wlpa} runs in $O(|E|\log|E|)$, re-optimizing on autoscaling,
service migration, or new partner integrations is an online operation
(milliseconds to low seconds at the scales we measured), not an offline
planning exercise. The data WLPA needs --- traffic volumes, compliance caps,
QKD hardware inventory --- typically already lives in existing telemetry.
Section~\ref{sec:guidance}'s two-condition checklist is deliberately written
so an operator can apply it to their own network before adopting coordinated
allocation at all: if lift cost is roughly constant and risk tracks traffic,
the current heuristic is provably fine and switching buys nothing.

\subsection{Limitations}\label{sec:limits}
The exploit-probability schedule, while grounded in a hybrid-argument
reduction and (for level~3) real hardware measurement, is still a modeling
choice, not a direct measurement of production compromise rates. The exact
ILP is intractable beyond a few hundred links; Algorithm~\ref{alg:wlpa} is
what runs at scale, with the path-coverage generalization (\S\ref{sec:pathcov})
handled only approximately. The real-network evaluation (\S\ref{sec:real})
uses a modeled, not measured, workload; capturing real per-link telemetry on
an enterprise-scale network is future work. The operator-loop integration
(Figure~\ref{fig:pipeline}) is conceptual: we did not build a service-mesh
policy-controller integration or measure enforcement latency on a live
system. RL/optimizer-comparison variance
across training seeds is not yet available and is flagged rather than
fabricated.

\section{Related work}\label{sec:related}
We organize by the four literatures a network-management reviewer would
check against, each stated as existing work, its limitation for this
problem, and what WLPA does differently.

\textbf{Network and cluster resource allocation.} Scheduling frameworks for
cloud/cluster resources~\cite{buyya2009,newman2015} allocate compute,
bandwidth, or placement to optimize throughput, cost, or availability.
\emph{Limitation:} none treat security posture as an allocatable resource
under a hardware-scarcity constraint --- their objective functions have no
notion of a scarce cryptographic upgrade or a weakest-link security floor.
\emph{WLPA's difference:} we pose posture assignment as the same class of
constrained-optimization problem this literature already solves for other
resources, with a closed-form optimum specific to the weakest-link
objective (Theorem~\ref{thm:ceiling}).

\textbf{SDN and network-management optimization.} Software-defined
networking separates the control plane from forwarding, enabling
centralized, programmatic policy
enforcement~\cite{mckeown2008,kreutz2015}, and this is precisely the control
plane WLPA's output targets (Section~\ref{sec:pipeline}). \emph{Limitation:}
this literature addresses \emph{how} to push a policy once decided (flow
rules, controller architecture), not \emph{which} posture to assign under a
QKD hardware budget --- the allocation decision itself is out of scope.
\emph{WLPA's difference:} we supply the missing decision layer --- an
optimal posture assignment --- that an SDN controller or service-mesh
policy engine would enforce; Section~\ref{sec:integration} details this
integration.

\textbf{QKD network resource allocation.} QKD-Alloc~\cite{bi2024} and
QKD-Route~\cite{kong2024} allocate QKD capacity by key-rate utility and
trusted-node routing respectively. \emph{Limitation:} both are
per-link/per-path rules in the sense of Section~\ref{sec:guidance} --- they
score and route around individual links or paths without a system-wide
weakest-link objective, so neither has a guarantee against the specific
failure mode Theorem~\ref{thm:sep} identifies. \emph{WLPA's difference:} we
evaluate both directly against WLPA on a real network
(Section~\ref{sec:real}, Table~\ref{tab:realent}) and show the gap is small
exactly where our own theory predicts it should be. Prior QKD network
surveys~\cite{mehic2020,dervisevic2024,wehner2018,cacciapuoti2020} motivate
QKD scarcity as a design constraint but do not address system-level
weakest-link optimization under budget.

\textbf{PQC migration and cryptographic agility.} Migration
guidance~\cite{mosca2018,joseph2022,ott2019,kampanakis2018} addresses
per-connection algorithm transition (which primitive, which parameter set,
how to phase the rollout). \emph{Limitation:} this guidance is silent on
\emph{system-level} allocation once multiple links compete for a scarce
stronger posture --- it tells an operator how to upgrade a link, not which
links to upgrade first under a budget. \emph{WLPA's difference:} we are a
system-level complement, taking per-link migration guidance as given
(Section~\ref{sec:background}'s posture levels) and adding the allocation
layer on top.

\textbf{Weakest-link security and attack graphs.} Attack-graph and
weakest-link security-economics
work~\cite{wang2008,sheyner2002,varian2004,anderson2006,grossklags2008}
establishes that system security is bounded by its weakest component.
\emph{Limitation:} this work is diagnostic (identify the weakest link) more
than prescriptive (optimally allocate a resource budget to raise it) under
an explicit hardware-scarcity model. \emph{WLPA's difference:} we
instantiate the weakest-link principle for the resource-scarce
QKD-allocation setting with a closed-form, efficiently-computable optimum
(Theorem~\ref{thm:ceiling}), not just a diagnostic metric.

\textbf{Zero Trust.} NIST SP 800-207~\cite{sp800207,kindervag2010,eo14028}
motivates per-link (never implicitly trusted) verification, consistent with
our per-link posture model; WLPA can be read as a resource-allocation layer
for a Zero-Trust deployment's cryptographic hardening budget.

\textbf{Concrete security.} Our exploit model is a standard hybrid-argument reduction to a
$(t,\varepsilon)$-security assumption~\cite{bellarerogaway2006,katzlindell2020,goldreich2001},
input to the framework rather than its contribution.

\section{Conclusion}\label{sec:conclusion}
Allocating scarce quantum-safe link postures is a network-management problem,
not only a cryptographic one. WLPA gives operators an exactly-optimal,
$O(|E|\log|E|)$ algorithm and a two-condition checklist for when their
existing traffic-ranked practice is already fine and when it silently fails.
Validated across 19 named experiments (Appendix~\ref{secA3}) including a real enterprise network, real
production traffic traces, and real QKD hardware, WLPA is designed as an
implementation-ready replacement for the ad hoc heuristics currently
embedded in network-management tooling's decision logic, with the
underlying theory --- a closed-form optimum, a proven separation dichotomy,
and an NP-hardness/approximation boundary for path coverage --- as its
foundation rather than its headline.

\section*{Declarations}
\bmhead{Funding} The authors declare that no funds, grants, or other support
were received during the preparation of this manuscript.
\bmhead{Competing interests} The authors have no competing interests to
declare that are relevant to the content of this article.
\bmhead{Availability of data and materials} The SNAP email-Eu-core dataset is
publicly available (\url{https://snap.stanford.edu/data/email-Eu-core.html}).
Implementation code for the four-architecture pipeline, the real-topology
experiment, and the post-quantum latency benchmark is provided as ancillary
files with this arXiv submission (directory \texttt{anc/code}), together with
\texttt{anc/RESULTS.md}; it will be deposited to a persistent public
repository upon acceptance.
\bmhead{Authors' contributions} B.~Gupta conceived the framework, implemented the algorithms and experiments, and drafted the manuscript. S.~Rana supervised the work, contributed to the problem formulation and analysis, and revised the manuscript. Both authors read and approved the final manuscript.
\bmhead{Acknowledgements} Not applicable.

\appendix
\section{Proof sketches}\label{secA1}
\noindent\textbf{Lemma~\ref{lem:hybrid}.} Standard hybrid argument: replace
each of the $N_e$ sessions on link $e$ one at a time with an ideal instance;
a distinguisher across $N_e$ hybrids implies a $t$-bounded distinguisher
against the underlying $(t,\varepsilon(\ell))$-secure primitive with
advantage $\ge1/N_e$ of the hybrid gap, giving the union bound
$N_e\varepsilon(\ell)$.

\noindent\textbf{Theorem~\ref{thm:ceiling}.} $\kappa$ bounds every link's cap;
$\sigma(B)$ and $\lambda(L)$ bound the level reachable by budget and latency
respectively, cap-aware. The minimum of the three is achievable by
construction (Algorithm~\ref{alg:wlpa}) and cannot be exceeded, since
exceeding any one term violates its respective constraint.

\noindent\textbf{Proposition~\ref{prop:noseparation}.} Under Hypothesis~H the
marginal cost of a level-2$\to$3 lift is link-independent, so the
breach-minimization objective is additively separable; its exact optimum is
an exchange-argument greedy solution that funds the largest per-unit breach
reduction, which under uniform risk is monotone in traffic $N_e$ --- exactly
the traffic-ranked rule.

\noindent\textbf{Theorem~\ref{thm:sep}.} Construct one cap-2 low-traffic link
$w$ and $k$ cap-3 high-traffic links $e_i$ with $N_{e_i}\gg N_w$; every
$\mathcal{R}_{\mathrm{PL}}$ rule, for every weight vector, scores each $e_i$
above $w$ (traffic dominance holds for any nonnegative weights), so a budget
sized to fund exactly the $e_i$ leaves $w$, the true system bottleneck,
unfunded: $S(Q)=1$ for every per-link rule versus $S^\star=2$ for coordinated
allocation. Scaling $k$ and the traffic gap scales the shortfall.

\noindent\textbf{Path-coverage NP-hardness.} Reduction from Set Cover: each
link's coverage of the paths it participates in corresponds to a set;
exactly covering all required paths within budget $B$ is Set Cover exactly.
The greedy algorithm inherits the standard $(1-1/e)$ submodular-coverage
approximation guarantee~\cite{nemhauser1978,feige1998}.

\section{Quantum circuit gallery}\label{secA2}
Ten Qiskit circuits generated for this work (QC1--QC10) --- covering QRNG
generation, BB84 (clean, eavesdropped, noisy, and full-session variants),
privacy amplification, the E91 entanglement-based alternative, and the WLPA
posture-encoding circuit shown conceptually in Figure~\ref{fig:pipeline} ---
are provided in full as ancillary files (\texttt{anc/quantum\_circuits\_full}); Figure~\ref{fig:qc6} in the main
text is representative.

\section{The 19 experiments, enumerated}\label{secA3}
For auditability, Table~\ref{tab:allexp} lists every named experiment
referenced in this paper by its identifier in the accompanying \texttt{RESULTS.md}
and code (ancillary files \texttt{anc/RESULTS.md} and \texttt{anc/code}), so the count in the abstract can be checked
directly against the supplement rather than taken on faith.
\begin{table}[htbp]\centering
\caption{All 19 named experiments referenced in this paper}\label{tab:allexp}
\begin{tabular}{@{}p{1.6cm}p{9.6cm}@{}}
\toprule
ID & Experiment \\
\midrule
Exp.\,1 & Algorithm correctness \& scalability sweep (Section~\ref{sec:ablscal}) \\
Exp.\,2 & Exact ILP match on tractable instances (Section~\ref{sec:exact}) \\
Exp.\,3 & Full baseline comparison, uniform risk (Section~\ref{sec:heuristics}) \\
Exp.\,4 & Adaptive-attacker budget sensitivity sweep (Section~\ref{sec:adaptive}) \\
Exp.\,5 & PQC latency hardware benchmark (Section~\ref{sec:pqcbench}) \\
Exp.\,6a & BB84 Qiskit simulation (Section~\ref{sec:pqcbench}) \\
Exp.\,6b & Real-hardware BB84, IBM \texttt{ibm\_marrakesh} (Section~\ref{sec:pqcbench}) \\
E1 & Base exploit-schedule sweep (Section~\ref{sec:defs}, Table~\ref{tab:extra}) \\
E2 & Heterogeneous QKD lift cost (Section~\ref{sec:guidance}, Table~\ref{tab:extra}) \\
E3 & Binding latency budget (Table~\ref{tab:extra}) \\
E4 & QKD-budget sensitivity sweep, breach probability vs.\ $B$ (Section~\ref{sec:adaptive}, Fig.~\ref{fig:budget}) \\
E5 & LP-relaxation certified optimality bound (Section~\ref{sec:exact}) \\
E6a & QKD key-rate feasibility envelope (Table~\ref{tab:extra}) \\
E7 & Real-workload robustness, Alibaba/Google traffic traces (Table~\ref{tab:extra}) \\
E8 & Cost-model (session-length) ablation (Table~\ref{tab:extra}) \\
E9 & Stackelberg (look-ahead) adversary (Table~\ref{tab:extra}) \\
E10 & Correlated/common-mode compromise, CVE-calibrated (Section~\ref{sec:extra}, Fig.~\ref{fig:corr}) \\
E11 & Bayesian uncertainty quantification (Table~\ref{tab:extra}) \\
E12 & Level-assignment honesty check (Table~\ref{tab:extra}) \\
\bottomrule
\end{tabular}
\end{table}
\FloatBarrier

\end{document}